\documentclass[preprint,groupedaddress]{revtex4-1}

\usepackage{hyperref}
\usepackage{epsfig}
\usepackage{graphicx}
\usepackage{subfigure}
\usepackage{latexsym}
\usepackage{color}
\usepackage{fullpage}
\usepackage{epstopdf}
\usepackage{dcolumn}
\usepackage{bm}
\usepackage{ulem}
\usepackage{units}

\begin{document}

\title{Unconventional magnetisation texture in graphene / cobalt hybrids}

\author{A. D. Vu$^{1,2}$, J. Coraux$^{1,2}$, G. Chen$^3$, A. T. N'Diaye$^4$, A. K. Schmid$^3$, N. Rougemaille$^{1,2,*}$}

\address{$^1$ CNRS, Inst NEEL, F-38000 Grenoble, France}
\address{$^2$ Univ. Grenoble Alpes, Inst NEEL, F-38000 Grenoble, France}
\address{$^3$ NCEM, Molecular Foundry, Lawrence Berkeley National Laboratory, Berkeley, California 94720, USA}
\address{$^4$ Advanced Light Source, Lawrence Berkeley National Laboratory, Berkeley, California 94720, USA}
\address{$^*$ nicolas.rougemaille@neel.cnrs.fr}

\date{\today}

\begin{abstract}
Magnetic domain structure and spin-dependent reflectivity measurements on cobalt thin films intercalated at the graphene / Ir(111) interface are investigated using spin-polarised low-energy electron microscopy. We find that graphene-covered cobalt films have surprising magnetic properties. Vectorial imaging of magnetic domains reveals an unusually gradual thickness-dependent spin reorientation transition, in which magnetisation rotates from out-of-the-film plane to the in-plane direction by less than 10$^\circ$ per cobalt monolayer. During this transition, cobalt films have a meandering spin texture, characterised by a complex, three-dimensional, wavy magnetisation pattern. In addition, spectroscopy measurements indicate that the electronic band structure of the unoccupied states is essentially spin-independent already a few electron-Volts above the vacuum level. These properties strikingly differ from those of pristine cobalt films and could open new prospects in surface magnetism.
\end{abstract}

\maketitle

\section*{Introduction}

Just a decade after being isolated, graphene has attracted considerable attention in the scientific community. Studies on this material allowed the discovery of rich and fascinating phenomena \cite{Novoselov2012, Grigorenko2012, Bonaccorso2010, Neto2009}, and opened new fundamental avenues to investigate purely two-dimensional systems. Moreover, because it shows exceptional mechanical \cite{Lee2008}, thermal \cite{Balandin2008}, electronic \cite{Du2008, Bolotin2008} and chemical \cite{Schedin2007} properties, 
graphene quickly emerged as a promising material for numerous applications. Potential technological use of graphene ranges from conductive electrodes \cite{Bae2010}, ultracapacitors \cite{Stoller2008}, radio frequency analog electronics \cite{Lin2010}, to sensors and solar cells \cite{Wang2008, Badhulika2015}, to name only a few. Combined with other materials, additional properties can even be induced in graphene. For example, graphene was made superconductive when contacted to Sn dots \cite{Allain2012}, semiconducting after grafting aryl groups \cite{Bekyarova2009} or magnetic when grown on top of a ferromagnetic metal \cite{Weser2010, Weser2011, Vita2014, Decker2013}.

With its weak spin-orbit coupling and hyperfine interaction, graphene also offers new opportunities in nanomagnetism \cite{Yazyev2010}. Graphene- and carbon-based hybrids are envisioned to play a key role in future spintronic devices and already opened a wealth of studies on spin transport \cite{Han2014} and spinterface phenomena\cite{Djeghloul2013, Djeghloul2015, Sanvito2010, Shick2014, Donati2013, Donati2014}. Although a spin-split band structure at the Fermi level could not be induced in graphene in contact with a ferromagnetic metal \cite{Rader2009}, a splitting was observed when graphene is deposited on a heavy metal \cite{Varykhalov2008}. In addition, graphene-based magnetic tunnel junction showed sizable tunnel magnetoresistance \cite{Decker2014}, low-resistance-area product combined with a high magnetoresistance \cite{Yazyev2009}, and other exciting spin-dependent effects have been observed and predicted (very long spin diffusion length \cite{Dlubak2012-1}, efficient spin-filtering \cite{Karpan2007}, etc).

If new properties can be induced in graphene by proximity effects, graphene can also modify the properties of its substrate. This is the case for example when graphene is in contact with a ferromagnetic metal. In particular, graphene has been shown to stabilise out-of-plane magnetisation in relatively thick Co films \cite{Rougemaille2012}. This result could be surprising at first sight since heavy transition metals with strong spin orbit interaction are often natural candidates to promote perpendicular magnetic anisotropy (PMA) in ferromagnetic thin films. Carbon being a light element, this observation suggests that orbital hybridisation at the graphene / Co interface might play an important role to promote PMA, similarly to what happens when C$_{60}$ molecules are deposited on top of a Co thin film \cite{Bairagi2015}. Graphene also serves in that case as an atomically-thick, efficient passivating layer, preventing oxidation of the ferromagnetic layer underneath \cite{Coraux2012, Dlubak2012-2}.

Here, spin-polarised low-energy electron microscopy is used to investigate in situ the magnetic domain structure of cobalt films grown on an iridium(111) surface or intercalated at a graphene / iridium(111) interface. Compared to pristine films, graphene-covered cobalt films exhibit unexpected properties. First, spin-dependent reflectivity measurements suggest that spin-scattering asymmetry is suppressed when the energy of the incoming electrons becomes larger than a certain threshold value, regardless the thickness of the cobalt film. Second, besides the fact that graphene favours perpendicular magnetic anisotropy, an unusually gradual thickness-dependent spin reorientation transition is observed. Analysis of the reorientation transition allows estimation of the surface anisotropy, which differs substantially from the value determined in pristine cobalt films. Third, measurement of all three Cartesian components of the magnetisation vector reveals that in-plane and out-of-plane magnetic domain patterns are characterised by different length scales that evolve differently when the cobalt thickness is increased. While in-plane domains show no change throughout the spin reorientation transition, meandering out-of-plane domains with well-defined, slowly decaying periodicity are observed when the cobalt film gets thicker. A complex three-dimensional spin texture is deduced from sets of magnetic images in graphene-covered cobalt films, in sharp contrast with what is found in pristine surfaces.

\section*{Experiment}

All measurements were done in situ, in the spin-polarised low-energy electron microscope (SPLEEM) available at the National Centre for Electron Microscopy of the Lawrence Berkeley National Lab. In a SPLEEM, an electron beam is directed at the sample in normal incidence and electrons that are backscattered from the surface are used for imaging. Because the energy of the incoming electrons is low, typically a few eV above the Fermi level of the material under investigation, the technique is surface sensitive and a few atomic layers are usually probed. The electron beam being spin-polarised, the technique is also sensitive to the surface magnetisation. The main strength of the technique is the capability to orient in any space direction the spin polarisation of the incident electron beam, thus allowing to probe unknown magnetic configurations. Moreover, since the energy of the spin-polarised electron beam can be tuned continuously, reflectivity measurements can provide useful information on the spin-split band structure of unoccupied electronic states \cite{Zdyb2002}. We refer the reader to review papers for more details on the technique \cite{Bauer2005-1, Bauer2005-2, Rougemaille2010}.

The samples studied in this work were fabricated and imaged under ultra-high vacuum conditions (base pressure in the $10^{-11}$ mbar range). An Ir(111) single crystal was used as a substrate and its surface was cleaned in a preparation chamber attached to the microscope, following a well-established procedure based on repeated cycles of Ar$^+$ ion sputtering and high temperature ($1200^\circ$C) flashes under oxygen ($10^{-8}$ mbar). A last temperature flash ($1200 ^\circ$C) was done in the microscope chamber to remove the oxide layer. Graphene was grown by chemical vapor deposition by exposing the Ir(111) surface to a pressure of ethylene (a few $10^{-8}$ mbar, typically for several tens of minutes), while keeping the substrate at about 600$^\circ$C. This relative low-temperature process produces a complete single-layer of graphene with a high density of defects \cite{Coraux2009}. Cobalt was deposited by molecular beam epitaxy at 300$^\circ$C and at a rate of about 0.3 monolayer (ML) per minute. Under such conditions, cobalt is intercalated at the graphene / Ir(111) interface in the form of a ultra-thin flat film and intermixing between Co and Ir can be neglected \cite{Drnec2015}.

Following previous works \cite{Rougemaille2012, Coraux2012, Vlaic2014}, we use the change of surface work function during Co deposition to track the  intercalation process, taking benefit from the large decrease of the vacuum level for most metal surfaces covered with graphene \cite{Giovannetti2008,Loginova2009,Vlaic2014, Murata2010}.

\section*{Results and discussion}

\subsection*{Suppression of spin-scattering asymmetry in graphene-coated Co ultrathin films}

\begin{figure}[ht]
\centering
\includegraphics[width=16cm]{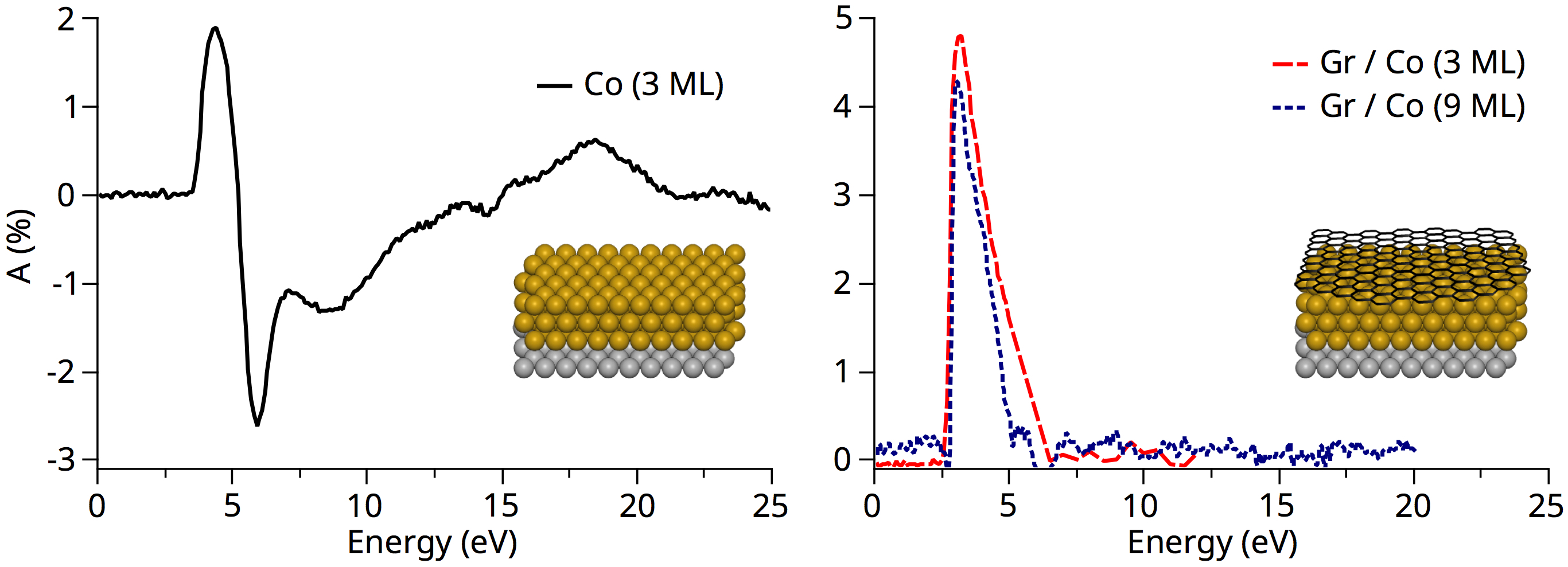}
\caption{Energy dependence of the spin asymmetry for a 3 ML-thick Co film grown on Ir(111) (left) and for a 3 and 9 ML-thick Co film intercalated between graphene and Ir (right). Energy is referred to the energy difference between the surface work function of the sample and the electron source.}
\label{fig:asymmetry}
\end{figure}

When a low-energy, spin-polarised electron beam is directed at a metal surface, part of the incident electrons enters the crystal and part of the incident intensity is backscattered. If the metal is ferromagnetic, the backscattered intensity depends on the relative orientation of the incoming spin polarisation and the direction of the local magnetisation within the material. This mechanism leads to a magnetic contrast in SPLEEM, which is used to image magnetic domain patterns. For common ferromagnetic metals, such as Fe, Co and Ni, the spin asymmetry $A$, defined as $A=(I_{+} - I_{-})/(I_{+} + I_{-})$, where $I_{\pm}$ is the reflected intensity for a spin polarisation $\pm P_{0}$, is a function of the incident energy. In most cases, the spin asymmetry changes in sign and amplitude depending on details of the unoccupied band structure of the material. Examples of energy-dependence of SPLEEM contrast can be found in previous literature \cite{Wu2005, Wu2006}. A typical spin asymmetry spectrum is reported in Fig.~\ref{fig:asymmetry} in the case of 3 Co monolayers (ML) deposited on Ir(111). The spin asymmetry is zero below the surface work function as the incident electron beam is totally reflected. The spin asymmetry first increases and then decreases to becomes negative before increasing again above 5 eV, typically. Except for specific energies where the spin asymmetry changes sign, it is non zero for energies ranging from 5 to 25 eV (and above).

When the same surface is graphene-terminated, the spin asymmetry is drastically modified. In particular, it shows a peculiar feature: while the spin asymmetry first increases as the incident energy becomes larger than the surface work function, it quickly decreases to becomes negligible above 5 eV, and up to at least 20 eV (Other measurements, not reported here, show that the spin asymmetry remains close to zero up to 35 eV). Within this energy window, the graphene-covered Co surface seems non magnetic. Although the Co surface is of course ferromagnetic, above a certain threshold energy, the incident spin-polarised electrons do not suffer substantial spin-dependent elastic or inelastic collisions. This suggests that the surface band structure of the graphene-covered Co film is not spin-split anymore above this threshold energy and/or that inelastic electron-electron collisions do not involve significantly electronic states close to the Fermi energy, where the density of state should be different for minority and majority spins. Considering that several atomic monolayers are usually probed in SPLEEM, even in ferromagnetic metals in which the inelastic mean free path is short \cite{Pappas1991, Getzlaff1993}, this indicates that the whole electronic band structure of the Co film is affected by the graphene topmost layer. Indeed, comparing the spin asymmetry of 3 and 9 ML-thick Co film intercalated between graphene and Ir does not show clear differences, suggesting that the effect extends through several monolayers.

This suppression of the spin asymmetry above a few electron-Volts is reminiscent of the unexplained 'breakdown effect' observed when ferromagnetic surfaces are covered with molecules \cite{Djeghloul2014}. After deposition onto Co surfaces of Co-substituted phthalocyanine (PcCo) molecules, both the amplitude and phase of the electron reflectivity were shown to become spin-independent for reflected primary electrons with an energy higher than 6 eV above the Co Fermi level. Although the physics governing this apparent suppression of surface magnetism in PcCo-coated Co thin films still remains unclear \cite{Djeghloul2014}, our results provide another instance of this breakdown effect, extending its generality to graphene-covered Co surfaces. 

\subsection*{Unusually gradual thickness-dependent spin-reorientation transition}

\begin{figure}[ht]
\centering
\includegraphics[width=16cm]{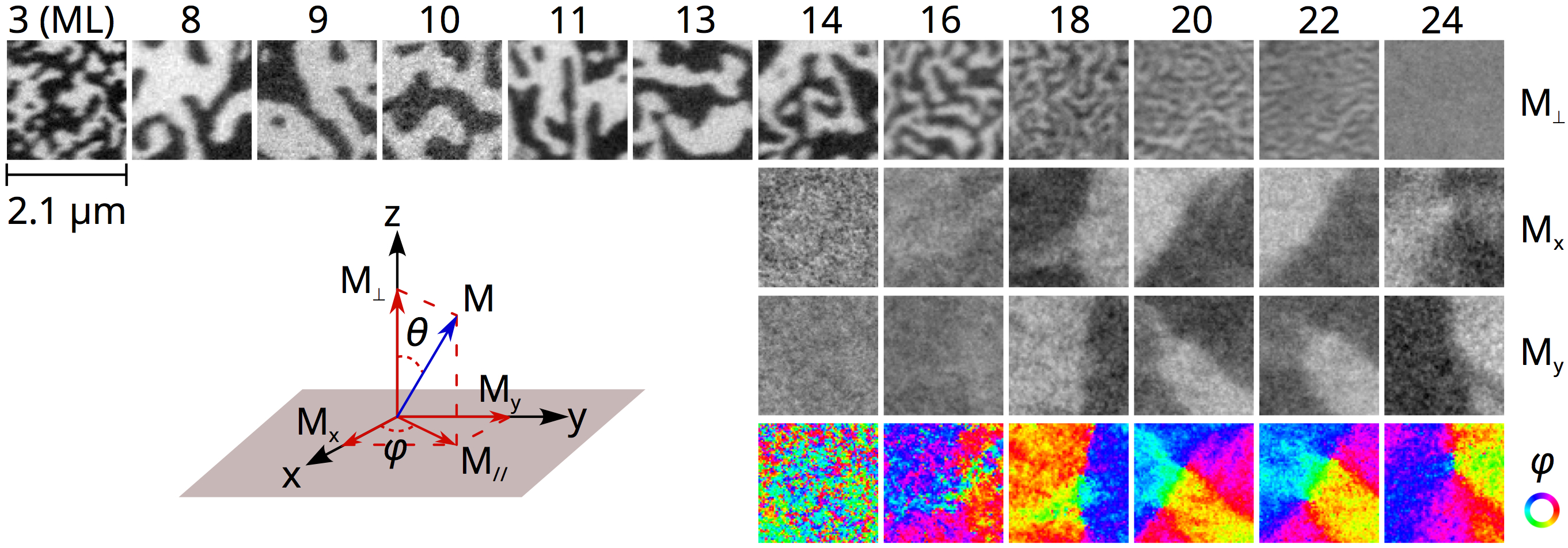}
\caption{SPLEEM images showing the out-of-plane (first row) and in-plane (second and third rows) domain patterns for three orthogonal incident spin polarisations as a function of Co film thickness in graphene / Co / Ir(111) heterostructures. The numbers indicate the thickness of the intercalated Co film in monolayers (ML). For Co films thinner than 13 ML, magnetisation is purely out-of-plane and no in-plane contrast is measured. For films thicker than 13 ML, magnetisation is in a canted state and SPLEEM contrast is observed for any incident spin polarisation. Above 24 ML, magnetisation has completely rotated into the film plane and no contrast is measured anymore when the incident spin polarisation is normal to the sample surface. The fourth row is deduced from the two orthogonal in-plane contrasts and gives the in-plane polar angle of the magnetisation using a colour code.}
\label{fig:SPLEEM} 
\end{figure}

In the following, we study the spin reorientation transition (SRT) that occurs when the amount of intercalated cobalt is increased, and we compare it to the SRT observed in pristine cobalt. For thin-enough films, magnetisation is out-of-plane, while it is in-plane for thicker films. To investigate how magnetisation rotates, we image magnetic domain patterns for different cobalt thicknesses in the three Cartesian coordinates. From the set of SPLEEM images, we deduce the local direction of the magnetisation vector as a function of Co thickness. SPLEEM measurements for Co films intercalated between graphene and Ir(111) are reported in Fig.~\ref{fig:SPLEEM}. When the Co film is thinner than 13 ML, magnetisation is purely out-of-plane, while it is purely in-plane for Co films thicker than 24 ML. In the intermediate thickness range, magnetisation is in a canted state and a magnetic contrast is measured for three orthogonal incident spin polarisations.

\begin{figure}[ht]
\centering
\includegraphics[width=8cm]{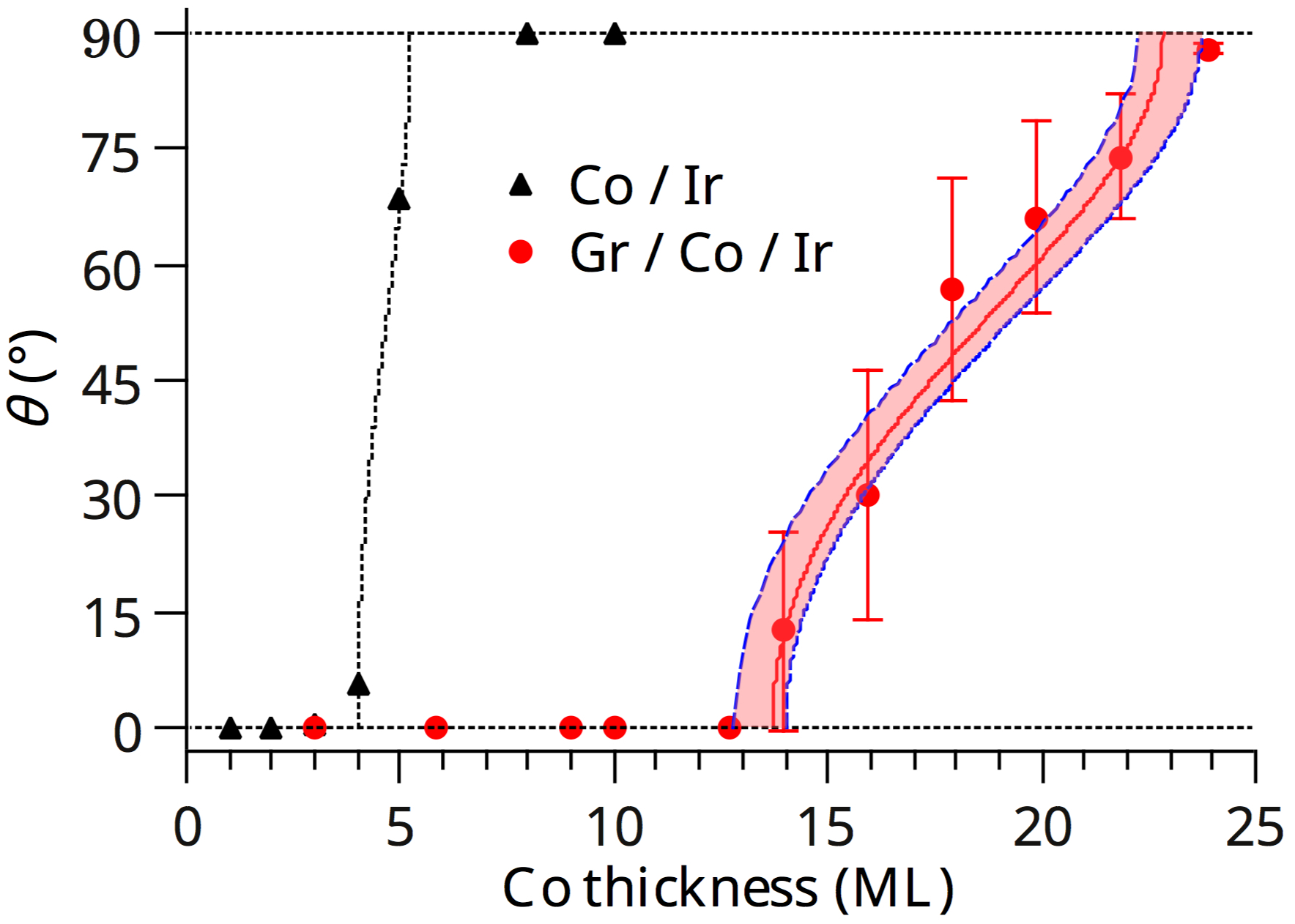}
\caption{Canting angle $\theta$ as a function of Co thickness in Co / Ir(111) and graphene / Co / Ir(111) heterostructures. The red curve shows the best fit, while the blue dotted curves illustrate how the fit varies when changing the $K_{2s}$ and $K_{4s}$ values (corresponding values in ~$mJ/m^2$ are respectively 2.06 and 0.20 for the upper blue curve and 2.25 and 0.19 for the lower one, the best fit giving $K_{2s}=2.20~mJ/m^2$ and  $K_{4s}=0.16~mJ/m^2$).}
\label{fig:canting} 
\end{figure}

The magnetisation vector can be represented in terms of the two spherical angles $\theta$ and $\varphi$, as illustrated in the inset of Fig.~\ref{fig:SPLEEM}. Across the SRT, $\mathbf{M}$ thus continuously rotates from $\theta = 0^\circ$ to $\theta = 90^\circ$. In the canted phase, the $\theta$ values can be extracted from Fig.~\ref{fig:SPLEEM} and plotted as a function of the Co film thickness (see Fig.~\ref{fig:canting}). To our knowledge, the thickness range over which the SRT occurs, about 10 ML, is unprecedentedly large compared to what has been reported in previous literature on different systems, such as Co on Pt and Pd \cite{Lee2002}, Ni on Cu(100) \cite{Klein2006, Klein2007}, Fe/Ni on Cu(100) \cite{Ramchal2004}, Fe on Ni/Pd(111) \cite{Yamamoto2010} and Co on Au \cite{Millev1996}. For comparison, the SRT in pristine Co films is also reported in Fig.~\ref{fig:canting} (black dotted curve). Like in many cases, the Co film shows in that case perpendicular magnetisation for thicknesses of a few atomic layers only ($\sim$ 4 ML) and is in-plane magnetised for thicker films (above $\sim$ 6 ML).

The SRT of our Co films can be described by the anisotropy flow model, in which the influence of both the surface and the interface on the magnetisation is accounted for by their respective phenomenological anisotropy constants \cite{Millev1996}. The phenomenological expression for the free-energy per volume unity reads:

\begin{equation}
 E = \widetilde{K_{2}} \text{sin}^2\theta + K_4 \text{sin}^4\theta
\end{equation}

where $\widetilde{K_2} = K_2 - \frac{1}{2} \mu_0 M_s^2$ contains the $2^{nd}$-order and the shape magnetic anisotropies, and $K_4$ is the $4^{th}$-order term of the magnetocrystalline anisotropy. Minimising $E$ with respect to $\theta$ gives three possible equilibrium states: $\theta = 0^\circ$, $\theta = 90^\circ$ and $\theta = \text{arcsin} \sqrt{-\widetilde{K_{2}}/2K_4}$. 
To explicitly consider the $\theta$ dependence of the Co film thickness $t$, we express $K_2$ and $K_4$ by the phenomenological ansatz: \cite{Millev1996}

\begin{equation}
K_2 = K_{2b} + \frac{K_{2s}}{t} \qquad K_4 = K_{4b} + \frac{K_{4s}}{t}\\
\end{equation}
Note that the interface anisotropy constants $K_{2s}$ and $K_{4s}$ are the sum of two terms associated with the Co / Ir(111) and graphene / Co interfaces. In this expression, $K_{2b}$ and $K_{4b}$ stand for the bulk anisotropy values. Hence, the SRT is delimited by two critical thicknesses:

\begin{equation}
\tilde{K_2} = 0 \Rightarrow t_{c1} = \frac{K_{2s}}{\frac{1}{2}\mu_0 M_s^2 - K_{2b}}\\
\end{equation}

\begin{equation}
\tilde{K_2} = -2K_4 \Rightarrow t_{c2} = \frac{K_{2s} + 2K_{4s}}{\frac{1}{2}\mu_0 M_s^2 - K_{2b} - 2K_{4b}}
\end{equation}

If $t_{c1} < t_{c2}$, the $\theta$ dependence of the Co thickness reads:
\begin{equation}
\theta = \text{arcsin} \left( \frac{1}{\sqrt{b + \frac{\Delta(1 - b)}{t - t_{c1}}}} \right)
\label{eq:theta_d}
\end{equation}
where $\Delta = t_{c2} - t_{c1}$ and $b = 2K_{4b}\big/(\frac{1}{2} \mu_0 M_s^2 - K_{2b})$.

Fitting our data with Eq.~\ref{eq:theta_d} (Fig.~\ref{fig:canting}), assuming the bulk values for Co at room temperature ($M_s = 1.44 \times 10^6 A/m$, $K_{2b} = 5.0 \times 10^5 J/m^3$ and $K_{4b} = 1.25 \times 10^5 J/m^3$) \cite{Oepen1997}, gives the surface anisotropy terms $K_{2s}$ and $K_{4s}$. For bare Co, we find $K_{2s}=0.64~mJ/m^2$ and  $K_{4s}= -0.03~mJ/m^2$, which are comparable to the values found for Co films on Au(111) or Pd(111) \cite{Oepen1997, Lee2002}. For graphene-covered Co, we obtain an unprecedentedly large $K_{2s}=2.20~mJ/m^2$ and  $K_{4s}=0.16~mJ/m^2$. Considering that $K_{2s} = K_{2s}^{Gr/Co} + K_{2s}^{Co/Ir}$ and taking $K_{2s}^{Co/Ir} = 0.8~mJ/m^2$ \cite{Broeder1991}, we estimate $K_{2s}^{Gr/Co} = 1.4~mJ/m^2$. 
The contribution of the graphene interface to the surface anisotropy is much larger than the value measured for the Co/Au(111) and Co/Pd(111) interfaces and is slightly smaller than the one reported for Co film on Pt(111)  \cite{Lee2002}. In addition to the remarkably large $2^{nd}$-order anisotropy term, the $4^{th}$-order term ($K_{4s}$) is found positive, while it is usually negative in the case of bare Co film in contact with Ir(111), Au(111), Pd(111) or Pt(111) \cite{Oepen1997, Lee2002}. Since $K_{4s}$ determines the upper thickness limit $t_{c2}$ of the SRT, its positive value in graphene-covered Co is associated to this exceptionally large thickness range.

\subsection*{A complex, wavy three-dimensional spin texture}

We note that in-plane and out-of-plane magnetic domains show distinct patterns: while in-plane domains are large, out-of-plane domains have meandering structures, similar to what is often observed in purely out-of-plane systems. Interestingly, these two patterns evolve differently as the thickness of the Co film is increased: in-plane domains are essentially thickness-independent, while out-of-plane domains are characterised by a periodicity that decreases as the Co thickness is increased (see Figure~\ref{fig:SPLEEM}). In-plane and out-of-plane domain patterns thus seem decoupled. This is another intriguing feature of the graphene / Co /Ir(111) system and could be the signature of an unusual spin texture.

\begin{figure}[ht]
\centering
\includegraphics[width=16cm]{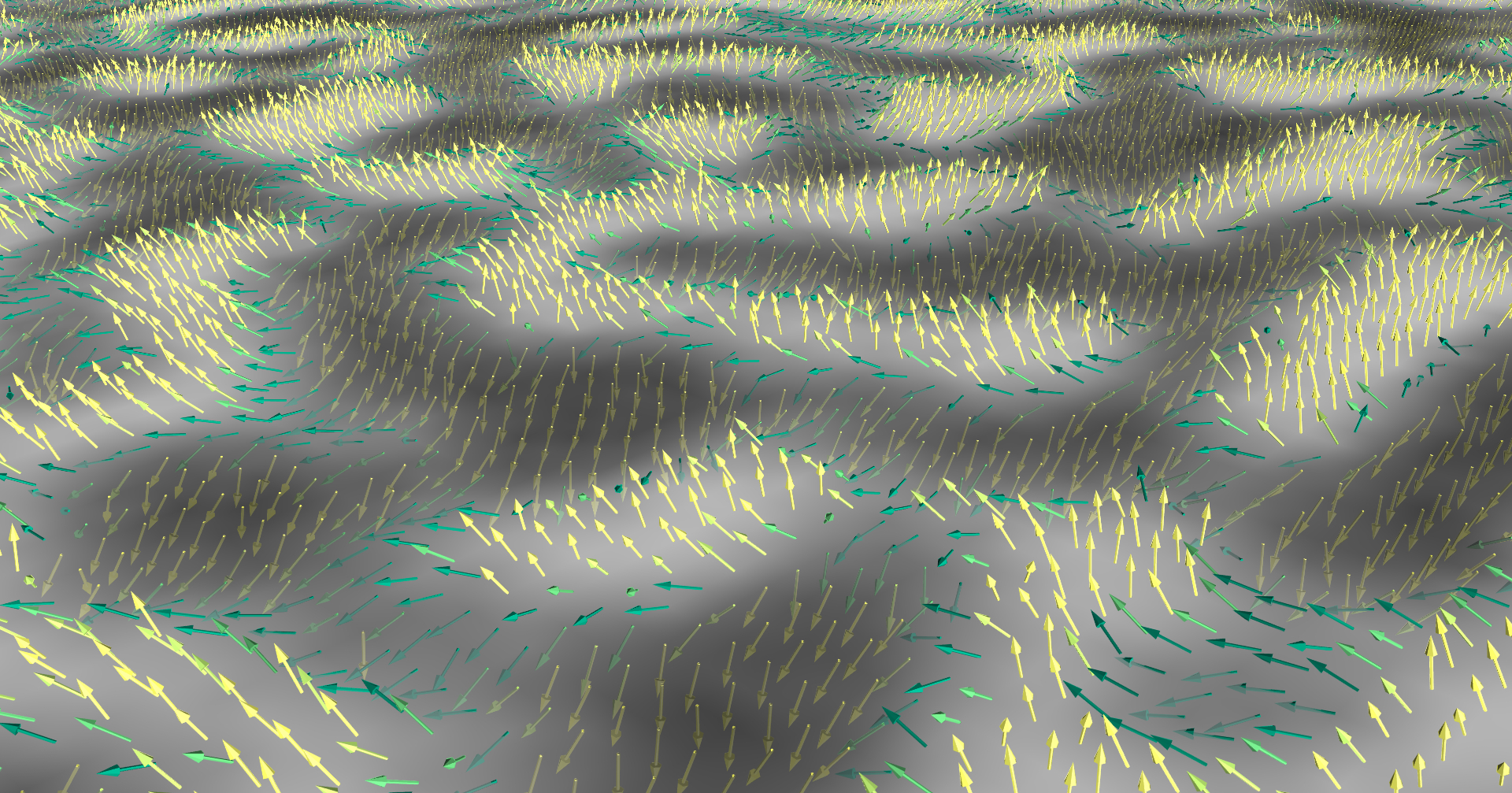}
\caption{Three-dimensional, pixel by pixel representation of the magnetisation vector (arrows) in the case of 16 ML of intercalated Co. Black and white contrasts give the out-of-plane component of the magnetisation, similarly to what is shown in Fig.~\ref{fig:SPLEEM}. Field of view is 3.2 $\times$ 2.4~$\mu m^2$.}
\label{fig:3D} 
\end{figure}

In the previous section, we determined the canting angle $\theta$ averaged over a set of SPLEEM images. In the following, we measure $\theta$, pixel by pixel, to get a local estimate of the canting angle and to determine how it evolves within a given field of view. Figure~\ref{fig:3D} shows such a pixel by pixel representation of the magnetisation vector in the case of 16 ML of intercalated Co. Black and white domains code for the out-of-plane component of the magnetisation vector, similar to what is shown in Fig.~\ref{fig:SPLEEM}. On top of these domains, the magnetisation vector is represented by coloured arrows (yellow when magnetisation is out-of-plane, green when it is in-plane). This figure indicates that magnetisation smoothly rotates in all space directions in form of waves across the field of view, similar to a wheat field blowing in the wind. Although the region between two neighbouring domains is of N\'eel-like type, a clear domain wall cannot be defined as the pattern resembles a flux closure configuration in three dimensions. The graphene / Co / Ir(111) system thus exhibits a unusual domain structure with a complex magnetisation texture, similar to what has been reported in other works \cite{Vedmedenko2002, Whitehead2008, Duden1996}.

We could wonder whether the magnetisation is uniform within the film thickness or if it also curls in the direction perpendicular to the surface. In fact, spin-twisted configuration may arise in some systems in which the spins at the outermost atomic layer feel a strong anisotropy and remain aligned along the surface normal, while those in the middle of the film are tilted towards the film plane to minimise the magnetostatic energy \cite{Thiaville1992,Popov2008}. Owing to its strong exchange interaction, such a spin-twisting is usually neglected in Co. However, the large anisotropy found experimentally could counterbalance the exchange interaction. Following previous works \cite{Thiaville1992,Popov2008}, we calculate the dimensionless parameter $\rho = K_{2s}/ \sqrt{A(\frac{1}{2} \pi M_s^2 - K_{2b})}$ characterising the spin-twisting state, where $A$ is the exchange stiffness constant. The limit case of uniform (non-twisted) magnetisation corresponds to $\rho \approx 0$ or $\text{tan}~(\rho) \big/ \rho \approx 1$. For the graphene / Co interface, $\rho \approx 0.28$ giving $\text{tan}~\rho \big/ \rho \approx 1.03$ (We assume bulk values for Co: $A=32~pJ/m$) \cite{Oepen1997}. Therefore, with good approximation, the magnetisation $\mathbf{M}$ can be assumed as uniform across the Co film.

\section*{Summary}

Using the capabilities of SPLEEM microscopy to probe the band structure of unoccupied electronic states and to resolve spin configurations in all space directions, we investigated the magnetic properties of in situ grown graphene / Co / Ir(111) heterostructures. Compared to bare Co / Ir(111) films, we found that the graphene capping layer drastically affects the spin texture and the spin-dependent band structure of the intercalated Co film. Similarly to what has been reported recently when PcCo molecules are deposited on a ferromagnetic surface \cite{Djeghloul2014}, the graphene layer seems to modify the spin-dependent band structure of the Co film: above an energy of a few eV typically, injecting spin-polarised hot electrons through a graphene / Co interface does not lead to a measurable spin-dependent reflectivity. Spin-scattering asymmetry is thus suppressed in a graphene-covered Co film, at least in a certain energy window. Moreover, studying the thickness-dependent spin reorientation transition in intercalated Co films, we observed an unusually gradual, continuous rotation of the magnetisation from the surface normal to in-plane. In fact, the SRT proceeds within a 10 monolayer window, meaning that, on average, the canting angle changes by of the order of only 10$^\circ$ per monolayer. From the critical thicknesses at which the SRT occurs, we determined the second and fourth-order anisotropy terms. The former is found to be unusually large, while the latter is found to be positive, contrary to what has been reported so far for Co. Analysing the magnetic domain patterns locally, we do not observe magnetic domains with uniform magnetisation separated by magnetic domain walls. Instead, we found a meandering, complex three-dimensional magnetic domain structure, in which the magnetisation vector is evolving in a wavy manner characterised by a short wavelength. The static and dynamic magnetic properties of graphene-covered Co films could open new prospects in nanomagnetism and surface magnetism.

\section*{Acknowledgements}

Experiments were performed at the Molecular Foundry, Lawrence Berkeley National Laboratory, supported by the Office of Science, Office of Basic Energy Sciences, Scientific User Facilities Division, of the U.S. Department of Energy under Contract No. DE-AC02-05CH11231. The authors thank the French National Research Agency for financial support via the ANR-12-BS-1000-401-NANOCELLS contract.

\section*{Author contributions statement}

All the authors contributed to the SPLEEM experiments and have discussed the results. A.D.V. analysed the data. A.D.V., J.C. and N.R. wrote the manuscript.

\section*{Additional information}

Competing financial interests: The authors declare no competing financial interests.

\end{document}